\documentclass[a4paper,12pt]{article}

\usepackage{amsmath}
\usepackage{geometry}
\usepackage{titlesec}
\usepackage{url}
\usepackage{graphicx}
\usepackage{hyperref}
\hypersetup{
    colorlinks=true,          
    linkcolor=blue,           
    citecolor=blue,           
    filecolor=blue,           
    urlcolor=blue             
}
\usepackage{float}
\usepackage{booktabs}
\usepackage{multirow}
\usepackage{color}
\usepackage{footmisc}

\titleformat{\section}{\normalfont\Large\bfseries}{\thesection}{1em}{}

\title{Time-Dependent Queuing Model for Traffic Congestion Using Mt/D/1/K: Simulation and Policy Insights}
\author{
  Jyoutir Raj\textsuperscript{\textdagger} \\
  \textsuperscript{\textdagger}School of Mathematics and Physics, The Queen's University of Belfast \\
}
\date{January 2025}

\begin{document}
\maketitle

\begin{abstract}
This study proposes a generalised macroscopic traffic simulation using a \(M_t/D/1/K\) queue to model congestion, using the Enniskillen to Belfast route as a case study. Empirical traffic data from Google's Directions API is used to calibrate the model, thus explaining peak commute times which we model using queue length. Simulations of staggered institutional start times showed significant reductions in queue lengths, suggesting time based interventions to improve rural to urban traffic flow.
\end{abstract}

\section{Introduction}
Road commuters often face congestion during peak periods, particularly around institutional start times such as schools and workplaces. This is especially true for regional routes like Enniskillen to Belfast, which pass multiple workplaces and urban areas, causing significant travel time delays. Understanding travel patterns and creating models to address this is a form of traffic management to reduce road congestion, aligning with the interests of Fermanah Community Transport of reducing travel times between regional routes. This project proposes a macroscopic traffic management simulation to model a queue along the Enniskillen to Belfast route, with the aim of explaining traffic congestion, suggesting simulated policies, and outlining future research directions.

Real time traffic data from Google's Directions API was collected over 30 days at 10 minute intervals to support the assumption that school and institutional start times are primary causes of congestion. This data allows for the construction of a Typical Day Model (TDM\footnote{The “Typical Day Model” is hereafter abbreviated as TDM.}), which explains the dynamics of daily travel times, including the peaks in the morning, midday and afternoon. Using the TDM, I assumed travel time is related to congestion which can be modelled by a queue, thus developing a nonstationary \((M_t/D/1/K)\) queuing model where the arrival rate \(\lambda(t)\) is a nonhomogeneous Poisson process.

The benefit of using these adjustable arrival rates is that the model can be tuned to specific traffic fluctuations, such as morning and evening peaks. The model represents macro traffic conditions, which evaluates overall system capacity rather than individual vehicle behaviour, ideal for proposing large-scale traffic interventions. Then, adjusting the model to simulate interventions like staggered institutional start times, I observed impacts on queue length and the consequent reduction of congestion and travel time.

Traffic flow is often modelled using macroscopic or microscopic approaches. Macroscopic methods, such as the Lighthill Whitham Richards (LWR) model \cite{Lighthill1955,Richards1956}, employ partial differential equations to describe density via conservation laws. While these model large-scale dynamics of traffic flows, building up and dissipation, they require detailed high-resolution data that may not always be available in practice. Microscopic or agent-based models \cite{Helbing2001} simulate individual vehicle interactions but demand significant computational resources and detailed behavioural inputs.

In recent years, queueing theory has been used as an alternative. Early research on finite capacity queueing networks (FCQNs) showed the impact of limited space on spillbacks and breakdowns \cite{SmithCheah1994}. However, few studies have explored the potential of the queueing theory framework to develop analytical traffic models, such as the M/G/C/C queue and the cell transmission model (CTM) \cite{Daganzo1994}, which utilised queueing theory for road traffic analysis.

My research builds on this subfield of queueing theory by proposing a time dependent queueing framework like \(M_t/D/1/K\) to provide a practical, data-first approach to represent daily congestion patterns, avoiding the complexities of PDEs or agent based simulations. By calibrating \(\lambda(t)\) using observed travel times from readily available databases, it approximates peak flows while requiring less granular data, making it particularly suitable to model land-based routes without the need of extensive data collection. In turn, this approach makes the model well suited to evaluate societal interventions, such as staggered start times, aimed at reducing queue lengths and improving travel reliability.

The findings from the model give insight into improving traffic flow and reducing congestion using the Enniskillen to Belfast route as a case study, but could be extended to any land based commute setting. This study contributes to regional transportation planning by demonstrating the applicability of nonstationary queuing models in rural to urban contexts and shows the potential benefits of time based traffic interventions targeting the problem of having too many vehicles on the road.

\section{Data Collection and Processing}

To accurately model traffic congestion on the Enniskillen to Belfast route, I collected empirical travel time data using Google's Directions API~\cite{google_api}. The data collection spanned a 30 day period from 25th October 2024 to 25th November 2024, during which travel times were recorded at 10 minute intervals over a 24 hour cycle. This discrete sampling gave a view of the variations in traffic patterns across a day, averaged over the span of a month, allowing me to spot differences between weekdays and weekends.

\subsection{Typical Day Model}

I processed the raw travel time data to develop the Typical Day Model (TDM). For each 10 minute interval, I calculated the mean travel time \(\bar{x}\) over the 30 day sample. To assess variability and ensure statistical confidence, I computed the 95\% confidence interval for the mean at each interval.

Treating each daily time segment as an individual sample helps account for variability in travel times per day and allows me to create the TDM. I use the TDM as a foundational reference for identifying peak congestion periods, which are essential to create the queuing model.

\begin{figure}[H]
    \centering
    \includegraphics[width=0.9\linewidth]{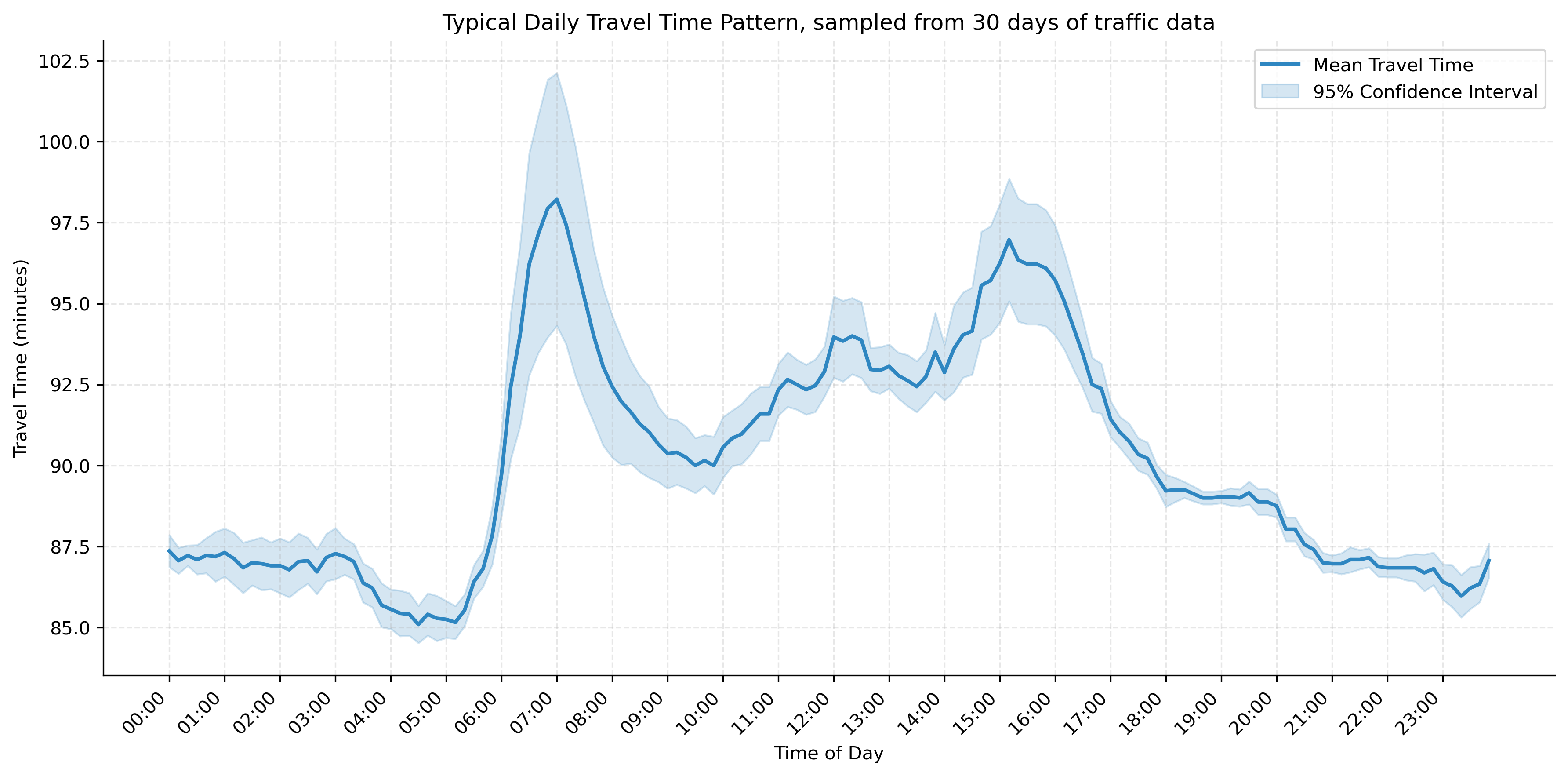}
    \caption{Typical daily travel time pattern with 95\% confidence intervals, based on 30 days of traffic data for the Enniskillen to Belfast route.}
    \label{fig:typical_day}
\end{figure}

The TDM in Figure~\ref{fig:typical_day} shows a baseline travel time regardless of time of day, with the journey taking 86 minutes even at 4am. Peaks are observed at 7–8AM and 15–16PM, supporting the hypothesis that institutional work start times significantly impact travel time.

\subsection{Insights from Data}

A closer analysis of the travel time data revealed clear differences between weekday and weekend patterns. To illustrate these differences, I categorised the travel times into three states using the 60th and 90th percentiles, denoted as \(Q_{0.6}\) and \(Q_{0.9}\), as threshold values:

\begin{itemize}
    \item \textit{Low travel times (green)}: \(t < Q_{0.6}\)
    \item \textit{Medium travel times (orange)}: \(Q_{0.6} \leq t \leq Q_{0.9}\)
    \item \textit{High travel times (red)}: \(t > Q_{0.9}\)
\end{itemize}

\begin{figure}[H]
    \centering
    \includegraphics[width=0.8\linewidth]{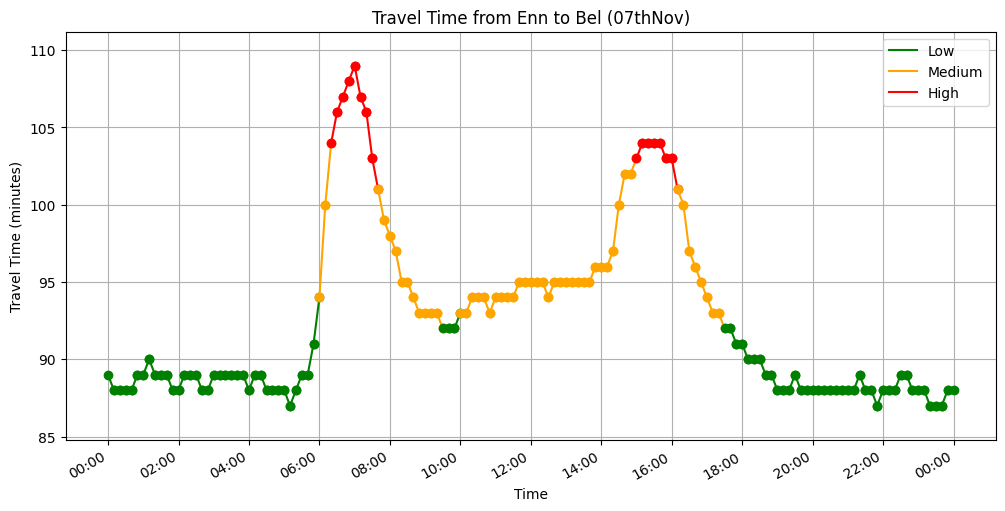}
    \caption{Travel time pattern for Thursday, 7 November 2024, on the Enniskillen to Belfast route, highlighting peak congestion periods during commuting hours.}
    \label{fig:weekday_pattern}
\end{figure}

\begin{figure}[H]
    \centering
    \includegraphics[width=0.8\linewidth]{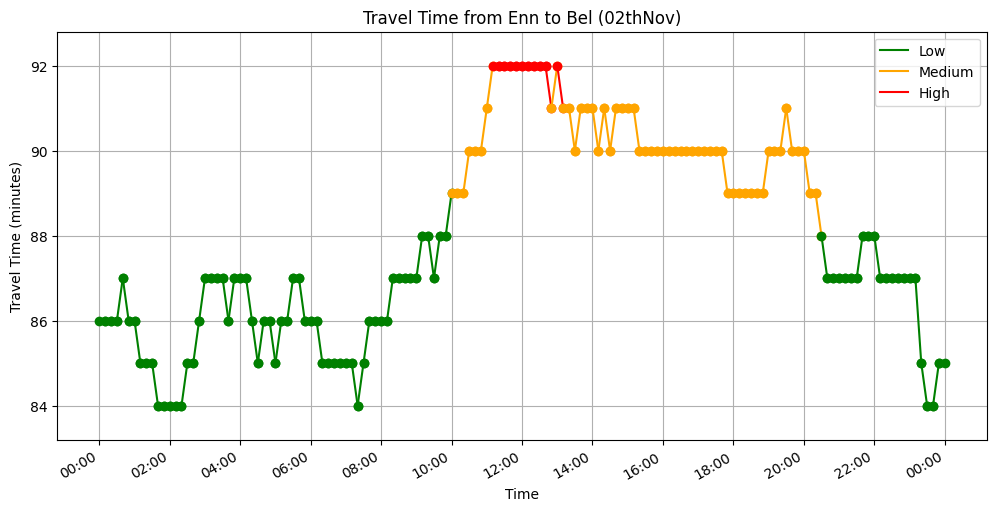}
    \caption{Travel time pattern for Saturday, 9 November 2024, on the Enniskillen to Belfast route, showing reduced maximum congestion and variability compared to weekdays.}
    \label{fig:weekend_pattern}
\end{figure}

I observed the following:

\begin{itemize}
    \item \textbf{Weekday Patterns}: Figure~\ref{fig:weekday_pattern} shows clear peaks in travel times during typical commuting hours (07:00--09:00 and 14:00--18:00), indicating higher traffic from school and work schedules.
    \item \textbf{Weekend Patterns}: Figure~\ref{fig:weekend_pattern} shows lower peaks (108 minutes on weekdays vs. 92 minutes on weekends) with reduced variability and no sharp spikes, indicating less commuting. The main commuting period shifts to midday (11:00--16:00).
\end{itemize}

Weekday peaks reach 108 minutes when schools and workplaces are in session, compared to weekend peaks of 92 minutes. Additionally, weekend traffic patterns are more flattened with no explicit peaks, unlike the pronounced peaks observed on weekdays. These figures show distinct traffic flow patterns when institutions are in session, supporting my hypothesis that they impact travel times.

\section{Model}

I model traffic along the Enniskillen to Belfast route as an $M_t/D/1/K$ queue, which has a nonhomogenous Poisson arrival process $\{\lambda(t)\}_{t \ge 0}$, a single server with constant service time $T=90.1$ minutes, obtained from mean travel-time from TDM, and a finite capacity $K$. Let $X(t)$ denote the number of vehicles in the system (on the route) at time $t$. Arrivals occur at rate $\lambda(t)$, estimated from TDM data; if $X(t)=K$, new arrivals are turned away. Because service is deterministic ($D$), each vehicle that enters at time $\tau$ departs exactly $T$ minutes later, provided it was not blocked at arrival. I assume the daily demand is roughly 120000 vehicles. Thus, the main dynamics arise from the time-varying Poisson arrivals $\lambda(t)$ (capturing peaks and troughs), the deterministic service time $T$, and the finite capacity constraint $K$.

\subsection{Assumptions and Limitations}
The rationale and key assumptions that guide my model include:

\begin{enumerate}
    \item \textbf{Time Varying Poisson Arrivals:} Vehicle arrivals follow a nonhomogeneous Poisson process with a time dependent rate \(\lambda(t)\). I model \(\lambda(t)\) as a Gaussian Mixture:
    \[
      \lambda(t) = \lambda_0 + \sum_{i=1}^{N} A_i \exp\Bigl(-\frac{(t - t_i)^2}{2\sigma_i^2}\Bigr),
    \]
    where:
    \begin{itemize}
      \item \(\lambda_0\) is the base arrival rate
      \item \(A_i\) is the amplitude of the \(i\)th peak
      \item \(t_i\) is the time of the \(i\)th peak
      \item \(\sigma_i\) is the duration of the \(i\)th peak
    \end{itemize}
    This approach allows calibrating \(\lambda(t)\) to the daily traffic variations observed in the TDM, and simulating policy changes like staggered start times.

    \item \textbf{Single and Finite Route Queue:} I treat the entire route as one queue, ignoring traffic lights and junctions. This is a macroscopic focus on overall congestion rather than local details. If the system already has \(K\) vehicles, arrivals are turned away until space becomes available.

    \item \textbf{Congestion Dependent Waiting Times:} Waiting time depends only on how many vehicles are already using the route. Factors like driver behaviour, weather, or accidents are not included.

    \item \textbf{Deterministic Service (D):} Each vehicle is assumed to travel for exactly 90.1 minutes (from the TDM) over the 82.4 mile route, corresponding to an average speed of about 55 mph. Vehicles do not move at a perfectly constant speed, but this simplification preserves realism enough for my macro level simulation.

    \item \textbf{Road Capacity (K)}: Assumed to be 20000 vehicles for this route, based on typical jam densities of 150 vehicles/km for a single lane 82.4 mile route \cite{Knoop2016}. This value serves as an upper bound for modelling.
\end{enumerate}

\section{Simulation}

Simulating this model and tuning the peaks in \(\lambda(t)\) based on empirical data from the TDM means I can align the model with observed school and work commute times. In the baseline scenario, \(\lambda(t)\) reflected existing traffic peaks between 07:00–09:00 and 14:00–18:00. Under the proposed staggered policy, arrival rates were adjusted to create staggered peaks, aiming to distribute traffic more evenly and reduce congestion during traditional peak hours.

\begin{figure}[H]
\centering
\includegraphics[width=\linewidth]{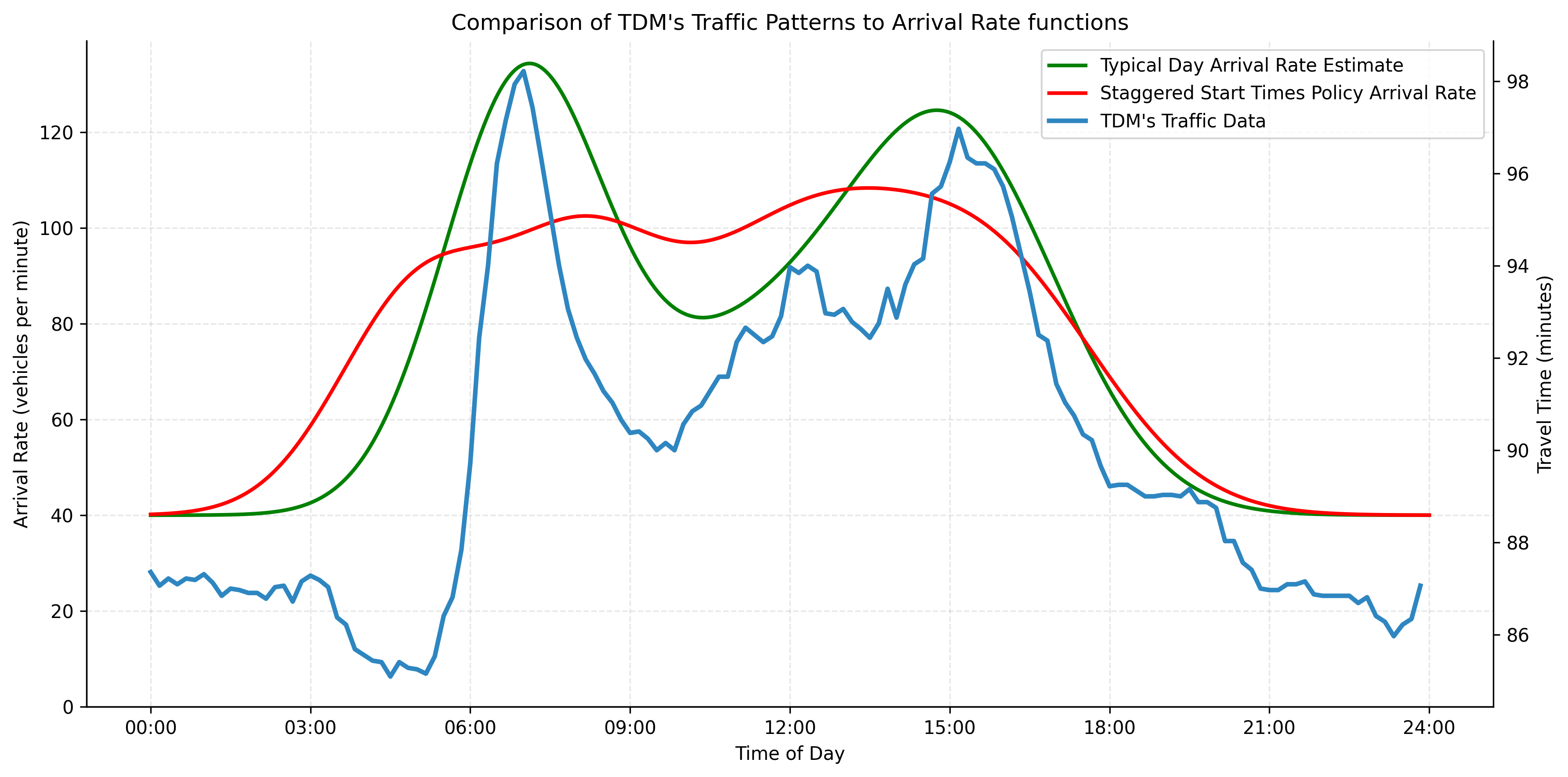}
\caption{Comparison of TDM's Travel Times to Arrival Rate Functions: The blue line is the TDM (observed travel times), while the green and red lines show the arrival rate estimates for the Typical Day and Staggered Start Times policies, respectively.}
\label{fig:traffic_patterns}
\end{figure}

\begin{table}[H]
\centering
\caption{Traffic Peaks for Typical Day and Staggered Start Times Models}
\resizebox{\textwidth}{!}{%
\begin{tabular}{cccccc}
\toprule
\textbf{Model} & \textbf{Peak} & \textbf{Description} & \textbf{Time (HH:MM)} & \textbf{Width (minutes)} & \textbf{Amplitude (veh/min)} \\
\midrule
\multirow{3}{*}{\shortstack{Typical\\Day}} 
& 1 & Morning Peak & 07:00 & 90 & 90 \\
& 2 & Mid morning Peak & 11:00 & 120 & 30 \\
& 3 & Afternoon Peak & 15:00 & 120 & 80 \\
\midrule
\multirow{5}{*}{\shortstack{Staggered\\Start\\Times Policy}} 
& 1 & Morning Peak 1 & 05:00 & 90 & 45 \\
& 2 & Morning Peak 2 & 08:00 & 90 & 45 \\
& 3 & Mid morning Peak & 11:00 & 120 & 30 \\
& 4 & Afternoon Peak 1 & 13:00 & 120 & 35 \\
& 5 & Afternoon Peak 2 & 16:00 & 120 & 45 \\
\bottomrule
\end{tabular}%
}
\label{tab:side_by_side_peaks}
\end{table}

Table \ref{tab:side_by_side_peaks} summarises the Gaussian peaks used in my arrival rate functions. Each peak's time, width, and amplitude are tuned to match the TDM's observed traffic trends. Figure \ref{fig:traffic_patterns} superimposes these arrival rate curves on the TDM data.

\begin{figure}[H]
\centering
\includegraphics[width=\linewidth]{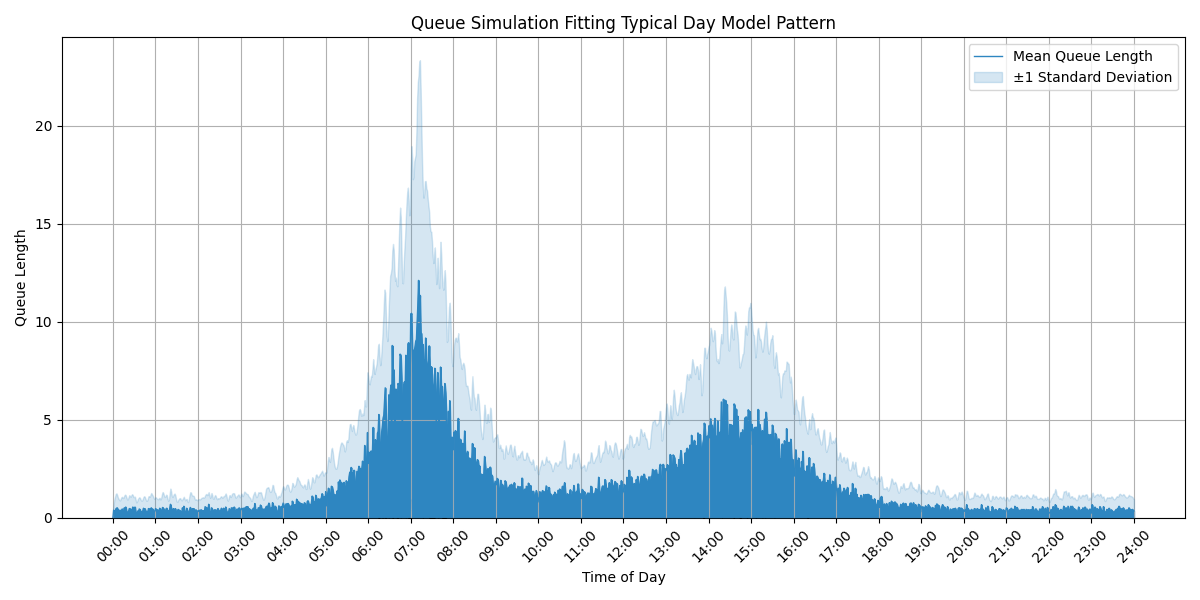}
\caption{Mean queue length over time under current traffic patterns for 50 simulations, showing pronounced peaks during traditional commuting hours.}
\label{fig:queue_typical}
\end{figure}

\begin{figure}[H]
\centering
\includegraphics[width=\linewidth]{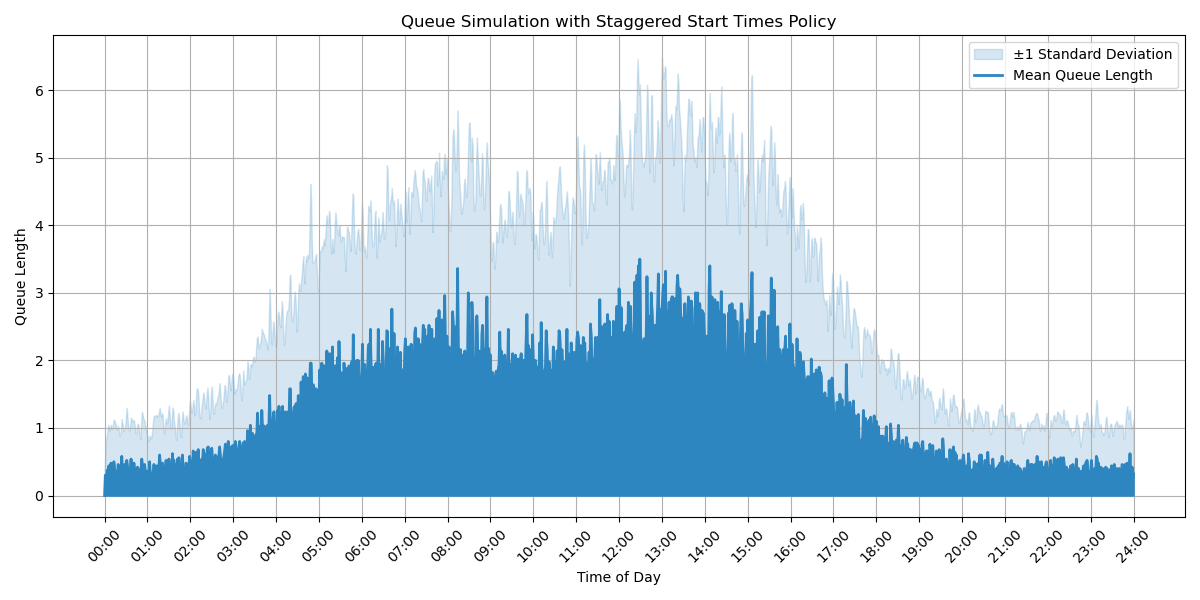}
\caption{Mean queue length over time under the proposed staggered policy for 50 simulations, showing reduced peak congestion and a more evenly distributed queue.}
\label{fig:queue_policy}
\end{figure}

Figure \ref{fig:queue_typical} shows significant increases in queue length during the morning (07:00–09:00) and afternoon (14:00–17:00) peaks, with a mean maximum queue length of 12.4 vehicles. Averaging over 50 simulations, the mean queue length is 2.84 vehicles (±1 standard deviation). The shaded region shows variability across runs.

Note that the model's output, queue length, is not identical to actual travel time. In queueing theory, one can approximate travel time by adding a deterministic service duration (the 90 minute route) to a waiting term:
\[
  \text{Travel Time} \approx
  \underbrace{90.1\,\text{min}}_{\text{Deterministic Service}}
  + \underbrace{\frac{Q(t)}{\lambda(t)}}_{\text{Waiting Time}},
\]
where \(Q(t)\) is the number of vehicles in the system at time \(t\) and \(\lambda(t)\) is the arrival rate. This follows from standard results like Little's Law \cite{Little1961}, which links queue length to delay. Strictly speaking, \(\frac{Q(t)}{\lambda(t)}\) is only approximate for nonstationary systems, but I adopt it as a heuristic to illustrate overall congestion trends.

\section{Conclusion}
The simulation suggests that staggered institutional start times reduce congestion by broadening arrival peaks, lowering simultaneous vehicle arrivals, and thus decreasing both queue lengths and overall travel times. Under the proposed policy, the mean maximum queue length drops from about 12.4 to 3.5 vehicles, while the average queue length decreases from 2.84 to 2.08 vehicles, roughly a 27 percent reduction, leading to a significant positive impact on the majority of daily commutes. Despite the simplicity of treating the entire route as a single queue, these findings align with the intuition that distributing vehicle departure times alleviates peak congestion.

A logical next step is to construct a more detailed microscopic model incorporating multiple bottlenecks (e.g. roundabouts, traffic lights) to test whether the staggering principle holds at finer scales. One could also explore alternative distributions for the time varying arrival rate \(\lambda(t)\) or refine and optimise the waiting time approximation to better match observed travel durations. Nonetheless, these results demonstrate that simple, time based interventions, such as staggered institutional schedules, can substantially improve traffic flow for most commuters.


\end{document}